# Autotuning and Self-Adaptability in Concurrency Libraries

Thomas Karcher
Institute for Program Structures and Data Organization
Karlsruhe Institute of Technology
76128 Karlsruhe, Germany
thomas.karcher@kit.edu

Christopher Guckes
Institute for Program Structures and Data Organization
Karlsruhe Institute of Technology
76128 Karlsruhe, Germany
ulcok@student.kit.edu

Walter F. Tichy
Institute for Program Structures and Data Organization
Karlsruhe Institute of Technology
76128 Karlsruhe, Germany
tichy@kit.edu

## ABSTRACT
Autotuning is an established technique for optimizing the performance of parallel applications. However, programmers must prepare applications for autotuning, which is tedious and error prone coding work. We demonstrate how applications become ready for autotuning with few or no modifications by extending *Threading Building Blocks (TBB)*, a library for parallel programming, with autotuning. The extended TBB library optimizes all application-independent tuning parameters fully automatically. We compare manual effort, autotuning overhead and performance gains on 17 examples. While some examples benefit only slightly, others speed up by 28% over standard TBB.

## 1. INTRODUCTION
Autotuning is a feedback-directed method that adapts a (parallel) program to a given hardware/software platform and input data characteristics, with the goal of optimizing one or more non-functional properties such as performance or energy consumption. Originally developed for numerical applications [5, 19], it is now applied to parallel software of all kinds [12, 13], including GPU applications [17]. However, so far it has been the programmer's task to prepare a program for tunability. In this paper, we demonstrate that it is possible to drastically simplify the work involved in this preparation. Our approach is to extend the concurrency library TBB for autotuning. When using this library, an application's source code requires no or very little preparation for tunability.

### 1.1 Autotuning
Autotuning is an iterative process illustrated in Fig. 1. It searches for an optimum by repeatedly executing and measuring an application under varying tuning parameter settings. Tuning parameters affect performance or energy consumption, but not correctness. Typical tuning parameters are thread count, work grain size, or pipeline stage replication. The measurements are taken on loops that take a significant proportion of execution time and whose performance is affected by the tuning parameters. The autotuner's job is to change tuning parameter values in an intelligent way so that the search for the optimum converges quickly.

Our autotuner works *on-line*. i.e. during production runs of the program. It repeatedly measures hotspots of the program under different parameter settings. The alternative, off-line tuning, is performed on benchmark data before pro-duction runs. Off-line tuning measures entire program runs, executed under different parameter settings. On-line tuning has the advantage that it can adapt to changing input characteristics, but has the disadvantage that tuning overhead occurs during production runs. Thus, in order to be useful, an on-line autotuner must find a performance improvement that amortizes the cost of tuning. It is imperative that an on-line tuner avoid testing too many bad choices.

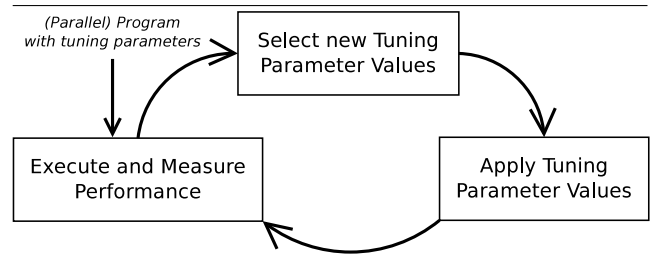

Figure 1 Autotuning loop. The autotuner receives execution time measurements from hot spots. Each time a measurement value arrives, the autotuner selects a new configuration of tuning parameters, applies this configuration to the program, and starts a new run.

The core of any autotuner is an efficient search algorithm. We employ the Nelder-Mead optimization algorithm [11] with slight modifications. The original version is designed for continuous and unlimited parameter spaces. Since tuning parameters are discrete and have bounds, we modified Nelder-Mead to round to nearest discrete points and to stay within given value ranges.

In our experience, autotuning excels if common assumptions about tuning parameters are violated. One common assumption is setting the thread count equal to the core count, which is often appropriate for CPU-bound applications. In non-CPU-bound applications, the best thread count is often harder to predict, so tuning will help in this case. Good values for other tuning parameters are also sometimes difficult to predict.

### 1.2 Threading Building Blocks (TBB)
Intel's TBB concurrency library provides an API for expressing high-level parallelism. Constructs for pipelines, data and task parallelism allow the programmer to feed parallel jobs to TBB's scheduler. TBB distributes the jobs among the cores and splits the jobs into smaller partitions if necessary. Load balancing by work stealing ensures short waiting times and avoids wasted CPU time.

Applications that employ TBB usually achieve acceptable parallel performance "out of the box", especially for CPU-bound applications. TBB's API forces the programmer to express an application's work in divisible jobs, e. g. a function body that is executed for each item in a data collection.

TBB already includes tuning parameters: The thread count for TBB's worker pool and the grain size for divisible jobs. The grain size specifies the amount of work per thread before contacting the scheduler for more work. The default value for the worker pool thread count is the core count on the system, while the default value for the grain size is 1000 (no unit given in the documentation; the number roughly corresponds to the size of the smallest task that a thread can get from the TBB scheduler).

## 2. PREPARING APPLICATIONS

An autotuner must be informed by the application about application-dependent tuning parameters and their value ranges. TBB provides generic tuning parameters. These can be tuned by generic code added to the library. If there are no application specific tuning parameters, the application source code does not need to be touched.

### 2.1 TBB Example Applications

We use the 17 examples shipped by Intel with TBB for evaluating our approach. Table 1 lists the examples and their properties. They are meant to demonstrate how TBB's API works and how to use TBB's constructs for data parallelism, pipelines, task management, and parallel data structures. Most of the examples focus on maximizing performance. An exception is LogicSim, a simulation of a logical circuit, which has fixed waiting times for certain circuit parts, thus simulating switching times. These waiting times dominate the run-time of the program.

### 2.2 Tuning Parameters

Applications usually aren't ready for autotuning. An on-line autotuner needs at least a single tuning parameter and a single measurement section. Applications that use TBB offer an elegant option: TBB's tuning parameters can be made accessible to the autotuner once and for all. The TBB concurrency constructs also contain sections that can be used for measurement. With our extended TBB, applications that use TBB become ready for on-line autotuning without touching their source code. Autotuning becomes truly automatic.

TBB provides an API with high-level concurrency constructs such as `parallel_for`, `parallel_reduce` or pipelines with filters. Most of these constructs take a collection of data items and a C++ functor as arguments. A TBB *partitioner* splits a collection into two. The TBB scheduler manages a pool of worker threads and assigns collections of appropriate size to workers, along with the functor. A worker invokes the functor for each data item in its collection. The scheduler decides whether a collection should be split because of load balancing. All of this happens under the hood of the concurrency constructs.

In the following, we concentrate on two tuning parameters: The number of worker threads and the amount of work per thread, called grain size. The default value for the number of worker threads is the number of cores and is determined upon application startup. This simple heuristic works well for CPU-bound jobs, but may be suboptimal for memory bound jobs or applications with I/O. Autotuning finds a good choice in all situations. Intel's recommendation for grain size ([8], section 3.2.1) is as follows: "Grainsize specifies the number of iterations for a 'reasonable size' chunk to deal out to a processor [...] A rule of thumb is that grainsize iterations of `operator()` should take at least 10,000–100,000 instructions to execute." Obviously, this recommendation is difficult to implement without some experimentation. Autotuning performs this experimentation automatically.

We explain the introduction of tuning parameters, the measurement loop, and how the autotuner works with the example named *Tachyon*. Tachyon is a ray tracing software. It calculates the color values of pixels in a 3D scene from a camera position, a light setting, and positions and sizes of objects and surfaces [15]. Tachyon uses TBB's `parallel_for` construct and thus uses the number of threads in TBB's worker pool and the grain size implicitly. The grain size influences how many pixels a single thread computes before it requests more work from the TBB scheduler.

The autotuning measurement loop in the example starts by invoking `parallel_for` and ends upon its return. With this simple approach, the autotuner receives a measurement after each complete 3D scene and thus advances one optimization iteration before the next scene starts to compute. A recalculation of a 3D scene happens each time the camera position changes, the light settings change, or objects move.

## 3. SPEEDUPS

With the number of worker threads and the grain size as tuning parameters within TBB, all programs receive free autotuning capability as long as they use a concurrency construct that include at least a single tuning parameter. We executed and measured the TBB examples in Tab. 1. First we explored the tuning potential without intelligent optimization, i. e. what are the worst and best tuning parameter values for a particular application. Afterwards we observe how the autotuner employs the Nelder-Mead algorithm to explore the search space and how well it performs.

We executed all experiments on a computer with 16 GB memory and two AMD Opteron 6168 processors with twelve cores each, clocked at 1.9 GHz.

### 3.1 Tuning Potential

Each tuning parameter is one dimension in a $k$-dimensional search space. Each point in that search space represents a particular tuning parameter configuration, i. e. a single value for each tuning parameter. The number of configurations per example is given in Tab. 1. By actually measuring the performance of each tuning parameter configuration, we get an exhaustive exploration of the search space. Many of the TBB examples contain two dimensions, one for the thread count and one for the grain size.

**Table 1** Intel ships these example programs with *Threading Building Blocks*. For the performance numbers, the number before the slash is the run-time, the second one the speedup. (LogicSim is not suited for performance optimization because of fixed chronological delays in simulated circuits.)

| Name | Purpose | Parallel Architecture | Tuning Parameters | Number of configurations | Sequential Time | Worst Case | Best Case | TBB time | Autotuned Ideal Time | Autotuned Average Time | Amortisation after Iteration |
|---|---|---|---|---|---|---|---|---|---|---|---|
| Bin Packing | Pack objects of different size in as few as possible bins | Flow graph | #threads, #packer threads (grain size) | 32x32 | 3.20 | 42.69 / 0.07 | 3.20 / 1 | 24.69 / 0.13 | 3.14 / 1.02 | 3.48 / 0.92 | 54 |
| Convex Hull | … of a collection of points | Data parallelism, parallel reduction | #threads, #points per task (grain size) | 32x4 | 10.16 | 10.16 / 1 | 0.07 / 146.13 | 0.09 / 111.53 | 0.62 / 16.34 | 0.88 / 11.48 | never |
| Count Strings | … in a text | Data parallelism | #threads, #characters per task (grain size) | 32x4 | 6.46 | 6.46 / 1 | 0.56 / 11.49 | 0.56 / 11.44 | 0.56 / 11.44 | 1.23 / 5.24 | never |
| Dining Philosophers | Simulation | Flow graph | #threads | 32 | 32.01 | 32.01 / 1 | 5.00 / 6.40 | 6.01 / 5.32 | 6.01 / 5.33 | 8.44 / 3.79 | 7046 |
| Fibonacci | 9 implementations | Data parallelism, divide-and-conquer | #threads, implementation selection | 32x9 | 0.036 | 1.67 / 0.02 | 0.03 / 1.29 | 0.53 / 0.07 | 0.04 / 0.8 | 0.32 / 0.11 | 7 |
| Fractal | Mandelbrot | Data parallelism | #threads | 32 | 9.62 | 9.62 / 1 | 0.42 / 22.98 | 0.42 / 22.75 | 0.42 / 22.88 | 2.26 / 4.26 | 3925 |
| LogicSim | Simulation of logic circuit | Flow graph | #threads | 32 | 10.01 | 10.01 / 1.00 | 10.01 / 1.00 | 10.01 / 1.00 | 10.01 / 1.00 | 10.01 / 1.00 | never |
| Parallel Preorder | Graph traversal | Data parallelism, task management | #threads | 32 | 12.99 | 12.99 / 1 | 0.74 / 17.60 | 0.77 / 16.88 | 0.75 / 17.23 | 2.20 / 5.91 | 826 |
| Polygon Overlay | Overlay of two polygon maps | Data parallelism | #threads, #points per task (grain size) | 32x10 | 0.56 | 6.16 / 0.09 | 0.07 / 8.17 | 1.60 / 0.35 | 0.11 / 5.01 | 0.91 / 0.61 | 1 |
| Primes | Sieve of Eratosthenes | Data parallelism, parallel reduction | #threads, numbers to test per task (grain size) | 32x4 | 12.72 | 12.72 / 1 | 0.54 / 23.66 | 0.54 / 23.43 | 0.54 / 23.56 | 1.56 / 8.16 | 3999 |
| Seismic | … wave simulation | Data parallelism | #threads | 32 | 0.06 | 0.06 / 1 | 0.01 / 8.90 | 0.01 / 7.05 | 0.01 / 5.67 | 0.01 / 3.78 | never |
| Shortpath | Shortest path with A* | Data parallelism | #threads, window size upon map generation | 32x3 | 3.67 | 3.67 / 1 | 2.16 / 1.70 | 3.03 / 1.21 | 3.08 / 1.19 | 3.02 / 1.21 | 3 |
| Pipeline Square | Calculate square numbers | Pipeline | #threads | 32 | 0.35 | 0.35 / 1 | 0.14 / 2.44 | 0.23 / 1.54 | 0.22 / 1.6 | 0.24 / 1.46 | 16 |
| Substring Finder | Find clusters of substrings | Data parallelism | #threads, grain size | 32x4 | 5.07 | 5.07 / 1 | 0.22 / 23.05 | 0.23 / 21.71 | 0.23 / 22.38 | 0.93 / 5.47 | 1493 |
| Sudoku | Calculate all solutions for a sudoku field | Data parallelism, task management | #threads | 32 | 15.90 | 15.90 / 1 | 0.67 / 23.85 | 0.67 / 23.84 | 0.067 / 23.75 | 2.19 / 7.25 | never |
| Tachyon | Raytracer | Data parallelism | #threads, #rays per task (grain size) | 32x10 | 3.41 | 3.43 / 0.99 | 0.14 / 23.77 | 0.19 / 17.73 | 0.15 / 22.60 | 0.28 / 12.11 | 67 |
| Tree Sum | Sum of all tree elements | Data parallelism, task management | #threads | 32 | 0.35 | 0.37 / 0.92 | 0.05 / 6.68 | 0.06 / 6.15 | 0.05 / 7.43 | 0.09 / 4.05 | 34 |

For each application, we measured sequential execution time, i. e. thread count set to 1, as a reference for speedups. We then extracted the worst and best configuration, i. e. the speedup that the worst and best choice for the tuning parameter values produce (see columns in Tab. 1). Between those two extremes lies the default configuration of TBB that every application runs with if the programmer doesn't change it.

We take a closer look at the Tachyon example. Thread count and grain size span a 2-dimensional search space. Every point in the search space represents a particular tuning parameter configuration. We let every example run 5 times with each configuration, measured the execution times and plotted the averages in Fig. 2. The single-threaded execution time, i. e. thread count set to 1, is 3.41 s – this is the reference point for speedups. The worst configuration achieves a speedup of 0.99 which suggests that the worst configuration is close to sequential execution. On the other hand, the best possible configuration runs with a speedup of 23.77. The standard TBB configuration already produces a speedup of 17.73 which leaves the autotuner room for improvement.

**Figure 2** Search space for the tachyon example (seismic wave simulation). The X and Y axes span the tuning parameters thread count and bin packer tasks, the Z axis shows the run-time. (The vertices mark the discrete configurations; although it is "illegal" to draw edges between vertices of a discrete space, we consider it illustrating. The black tiles represent areas of the search space that the autotuner explored.)

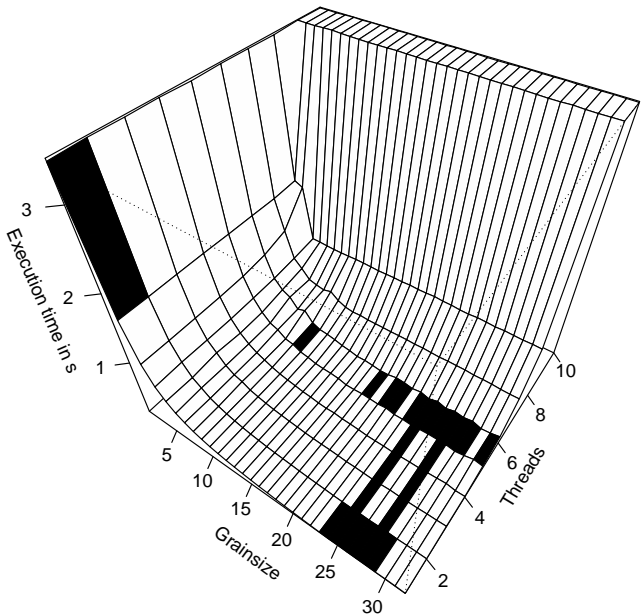

Overall results vary drastically: While e. g. Tachyon seems scalable but also improvable by autotuning, bin packing seems to suffer from even considering parallelism. With actual speed-down from sequential run-time with TBB default values, the autotuner can rescue some of the performance by reducing the thread count. Also, the corridor between the TBB standard configuration and the best configuration is fluctuating from example to example: Programs such as Convex Hull, Fractal, and Primes execute already near-optimal without autotuning while others such as Fibonacci, Polygon Overlay, and Shortpath can profit drastically by deviating from the TBB standard configuration.

## 3.2 Performance Improvements

The autotuner's task is finding the optimal tuning parameter configuration – or at least some configuration in the neighborhood of it. Our variant of the Nelder-Mead algorithm works on tuning parameter values, i. e. bounded intervals of integers, in contrast to the original continuous version of the algorithm. When executed repeatedly, we start each time with a simplex of fresh randomly chosen points in the search space to avoid getting stuck in the same local minimum. With repeated execution, the autotuner may arrive at different final configurations because of fluctuating system behavior and the random initialization phase in our variant of the Nelder-Mead algorithm.

We let each example run 15 times with the autotuner; the column "Autotuned Ideal Time" in Tab. 1 shows the speedup of the best configuration that the autotuner ever reached in any run. On average, the autotuner reached configurations that result in the speedup given in column "Autotuned Average Time".

Autotuning introduces overhead by itself and possibly slow-down because it lets the application run with bad tuning parameter configurations. The column "Amortization after Iteration" indicates the average number of iterations a program executes before it gains actual speedup due to autotuning. Some applications leave little or no room for speedup between the TBB standard configuration and the best configuration but still experience slow-down for bad tuning parameter values. With those applications, amortizing is hard to achieve, e. g. Count Strings, Seismic, and Sudoku. The value "never" indicates either that it is impossible to amortize or that the application ran too short to reach it.

In the Tachyon example, the autotuner doesn't know anything about the search space (Fig. 2) at first. It takes the autotuner 67 iterations to explore the search space sufficiently to converge to the configuration it considers best and to amortize the overhead it introduced. Note that the final configuration found by the autotuner is not guaranteed to be the best possible configuration in the whole search space. In the case of Tachyon, the final autotuned configuration varies heavily from run to run: There are runs where the autotuned version gets ahead of the TBB version (speedup 22.6 vs. 17.7) while there are also runs where the autotuned version falls behind with an average speedup of 12.11.

## 4. RELATED WORK

Performance optimization is a vast field with many solutions. Autotuning is a part of that field and comes in several flavors: (1) On- and offline, (2) domain-specific and domain-oblivious, i. e. tailored for a particular application or designed for generic utilization. In all cases, a programmer usually needs to invest time and effort to get an application ready for autotuning: Exposing tuning parameters, finding a measurement loop, and attaching the autotuner.

*ATLAS* [19, 20] established domain-specific autotuning for matrix multiplications in 1997. At installation time, microbenchmarks measure hardware properties such as cache and RAM access times. The results help decide which implementation variant to choose from and configuring tuning parameter values. Autotuning matrix multiplication has been popular ever since [4, 7, 18].

*Active harmony* [5] optimizes distributed applications. It supports domain-oblivious off- and on-line autotuning. Active harmony consists of several modules, among them a module for choosing the best performing library out of a collection and a tuning daemon that is supposed to run on a dedicated node. Most of the glue between source code and tuning daemon is specified in a specially designed language in a separate file. The tuning daemon uses the *parallel rank ordering (PRO)* algorithm [16] to optimize the application on-line with rapid convergence. PRO is a distributed descendant of the Nelder-Mead search algorithm [11].

*FIBER* [10] is an off-line autotuning framework targeted at distributed numerical applications. The programmer exposes tuning parameters via annotations in the source code. These annotations feed FIBER numerical knowledge about tuning parameters and measurement loops. FIBER uses that knowlegde to optimize the numerical application.

*MATE* [3] is an autotuning solution that focuses on distributed MPI applications. MATE uses two operational modes: (1) implicitly tuning MPI-intrinsic parameters such as buffer sizes for communication and (2) explicitly tuning application-specific tuning parameters that the programmer exposes as such. Mode (1) works for every MPI application without modifying the application's source code while mode (2) requires programmer intervention.

*Atune* [13, 14] uses parallel design patterns for providing the programmer with frequently used parallel building blocks. The programmer implicitly uses tuning parameters and measurement loops that are embedded in those building blocks, thus enabling the application for autotuning without noticing. The off-line autotuner extracts the tuning parameters, reduces the search space in a preparation step and performs search-based off-line optimization on the application afterwards. Atune needs the design patterns to be specified in a domain-specific configuration language called *TADL*.

*XJava* [12] is a Java dialect that provides flexible pipeline constructs. These constructs allow the programmer to express many types of high-level concurrency patterns such as master-worker, worker pool, or pipeline. Nested constructs are possible. XJava's source code transformer takes XJava code and outputs regular Java source code. During translation, XJava recognizes the concurrency patterns and introduces appropriate tuning parameters and measurement loops.

Autotuning with GPU-specific tuning parameters [17] shows promising results. Tillmann et al. exposed common GPU tuning parameters and attached a domain-oblivious autotuner to GPU applications.

*PetaBricks* [1, 2] provides a language for expressing algorithmic variants. The PetaBricks autotuner measures each variant unter laboratory conditions and saves a profile. At run-time, the autotuner decides based on that profile which variant to choose.

There are frameworks for algorithmic skeletons that provide high-level parallelization constructs – similar to those of TBB – and generate customized program variants for parallel CPUs or GPUs, e.g. SkePU [6].

Compilers are a mightier tool to introduce or extract tuning parameters, as well as employing more complex program transformations than changing a value of a tuning parameter. *INSIEME* [9] is an example for a compiler with focus on optimization. Contrary to the majority of autotuners, *INSIEME* delivers a pareto front of best-candidate configurations for multiple objectives instead of one best configuration.

## 5. CONCLUSIONS
We demonstrated how an application can become autotunable without programmer intervention. By exposing tuning parameters in Threading Building Blocks to an autotuner, applications that use this library tune themselves. Our approach can be complemented with application-specific tuning parameters, such as alternative algorithms or loop tilings. The effect of autotuning depends on the application: Some examples such as Bin Packing and Tachyon improve significantly, with up to 28% increase in speedup over the TBB default configuration. Others change their performance only slightly. There are even programs that slow down somewhat, because they cannot be accelerated with autotuning, while the on-line tuning process adds overhead, in particular by exploring alternatives that run slowly.

The tuning behaviors we observed suggest that a similar approach may work with other concurrency libraries. Future work will explore the behavior of other autotuners and their performance improvements.